\documentclass[preprint,aps,prc,showpacs,showkeys]{revtex4}

\usepackage{graphicx}
\usepackage{bm}

\begin{document}
 
\title{ Resonance lifetime in Boltzmann-Uehling-Uhlenbeck theory: 
        observable consequences }
 
\author{ A.B. Larionov \footnote{On leave from RRC "I.V. Kurchatov Institute",
         123182 Moscow, Russia}, 
         M. Effenberger \footnote{Present address: SAP AG, 
         D-69190 Walldorf, Germany}, 
         S. Leupold, U. Mosel }

\affiliation{ Institut f\"ur Theoretische Physik, Universit\"at Giessen,
          D-35392 Giessen, Germany }

\date{\today}

\begin{abstract}
Within the transport BUU theory we study the influence of the choice 
for the $\Delta$-resonance lifetime on pion-nucleus reactions and on  
heavy-ion collisions at 1 A GeV. A quite small effect of these 
modifications on the pion absorption on nuclei is found provided that the 
absorption probability of the $\Delta$-resonance is modified consistently.
Observable effects in the case of heavy-ion collisions are demonstrated.
\end{abstract}



\maketitle

\section{ Introduction }

It is usually assumed that the lifetime of a resonance is given by
its inverse total width $1/\Gamma$. This assumption is widely
used in the transport simulations of nuclear collisions in order to
describe the decays of various baryonic and mesonic resonances.
However, such a picture is not always correct.

Let us consider a resonance as the intermediate state of the two-body 
(e.g. $\pi N$) scattering. The lifetime of the resonance depends crucially 
on the relation between its width $\Gamma$ and the energy spread of the
incoming particles $\Delta E$ \cite{Joachain}:
(i) For $\Delta E \ll \Gamma$ (broad resonance) the lifetime
is given by the derivative of the phase shift with respect 
to the center-of-mass (c.m.) energy:
\begin{equation}
      \tau = { d \delta(E_{c.m.}) \over d E_{c.m.} }~,
                                                             \label{taudel}
\end{equation}
which is the time delay in the transmission of the scattered wave 
in the case of scattering with only one partial wave \cite{Mes,DP96}.
(ii) In the opposite limit $\Delta E \gg \Gamma$ (narrow resonance)
the averaging of Eq.(\ref{taudel}) can be done over $E_{c.m.}$ weighted
with the cross section which leads to the average life
time of $1/\Gamma$ \cite{DP96}.

In the Boltzmann-Uehling-Uhlenbeck (BUU) transport theory the colliding 
particles -- by definition -- have fixed energies and momenta. 
Therefore, the {\it correct} lifetime of a resonance in BUU is that of 
Eq.(\ref{taudel}) (see also Ref. \cite{David99}). In particular, the life 
time of the $\Delta(1232)$ resonance is given by Eq.(\ref{taudel}) with
$\delta(E_{c.m.}) \equiv \delta_{33}(E_{c.m.})$, where $\delta_{33}$
is the phase shift of the $\pi$N scattering in the S=3/2, I=3/2 channel. 

The two definitions of the lifetime, the {\it wrong} one as $1/\Gamma$ 
and the {\it correct} one as the phase shift derivative, have a completely 
different dependence on the c.m. energy (see Fig. 1 and Eqs. 
(\ref{taulife}),(\ref{gamma1}) below): the inverse width decreases 
monotonically with $E_{c.m.}$ while the phase shift derivative 
has a maximum near $m_\Delta=1.232$ GeV. Especially, at threshold 
$E_{c.m.}=m_N+m_\pi=1.076$ GeV, the inverse width becomes infinitely large,
while the phase shift derivative is zero. In view of the fact that most
transport simulations use the wrong $1/\Gamma$ prescription it is interesting 
to find observable signals sensitive to the different lifetime prescriptions. 
In the present work -- using the BUU model -- we study the influence of the 
$\Delta$ resonance lifetime on the pion absorption on nuclei and on some 
pionic observables from heavy-ion collisions at 1 A GeV. Note that a quite
thorough investigation of the analogous effects within BUU in the Fermi
energy domain has been done by Morawetz and coauthors. Their works
concern the nonlocal and quasiparticle corrections to the nucleon-nucleon
scattering (see Ref. \cite{Mor01} and references therein).

The paper is structured as follows: In Sect. II, the modifications of
the $\Delta$ resonance absorption and rescattering probabilities in
nuclear medium are discussed, the parameterizations of the $\Delta$ 
width and lifetime for transport simulations are chosen. Sect. III 
contains numerical BUU results. Coclusions are given in Sect. IV.

\section{ In-medium $\Delta$ lifetime } 

Recently Eq. (\ref{taudel}) has been generalized in Ref. 
\cite{leupold01} to the case of the decay of the 
$\Delta$ resonance with multiple final channels, i.e. when the 
total width $\Gamma$ contains not only the $\Delta \to N \pi$ contribution,
but also the absorptional contributions like, e.g. $\Delta \to N^{-1} N N$.
The result of Ref. \cite{leupold01} for the decay time
of a uniform bunch of resonances which do not interact with each other   
(see Eq. (3.3) in \cite{leupold01}) rewritten here for 
simplicity in the nonrelativistic approximation is as follows  
\begin{equation}
        \tau = {1 \over 2} A (1-K)~,
                                                            \label{taulife}
\end{equation}
where 
\begin{equation}
           A(\omega,{\bf p})
         = {\Gamma(\omega,{\bf p}) \over 
            (\omega-{\bf p}^2/2m-\mbox{Re}\,\Sigma^+(\omega,{\bf p}))^2 
          + \Gamma^2(\omega,{\bf p})/4}     
                                                            \label{spfun}
\end{equation}
is the spectral function and
\begin{equation}
         K \equiv { \partial \mbox{Re}\,\Sigma^+(\omega,{\bf p})
               \over \partial \omega }
            + { \omega-{\bf p}^2/2m-\mbox{Re}\,\Sigma^+(\omega,{\bf p})
                \over \Gamma(\omega,{\bf p}) }
              { \partial \Gamma(\omega,{\bf p}) 
                 \over \partial \omega }
                                                           \label{K}
\end{equation}
with the total width $\Gamma(\omega,{\bf p}) = 
-2\mbox{Im}\,\Sigma^+(\omega,{\bf p})$. In the case of a resonance gas 
in vacuum, i.e. when only the $\Delta \to N \pi$ decay channel is open, 
Eqs. (\ref{taudel}) and (\ref{taulife}) are equivalent. This can be
easily seen by introducing the phase shift (c.f. \cite{Mes})
\begin{equation}
       \delta 
           = \arctan{ \Gamma(E_{c.m.}) 
                              \over 
                    2(E_R(E_{c.m.})-E_{c.m.}) }~,  
                                                         \label{phshift}
\end{equation}
with
$E_R = m + {\rm Re}\,\Sigma^+(E_{c.m.})$,
$E_{c.m.} = m + \omega - {\bf p}^2/2m$.

In transport simulations one deals with the partial lifetimes 
$\tau_i$ of resonances with respect to decay into different channels, 
including absorption and rescattering. If one uses e.g. 
$1/\Gamma_{\Delta \to N \pi}$ as the lifetime for the $\Delta$ 
with respect to the $N \pi$ decay channel, this corresponds to the use 
of the ``standard'' cross section $d \sigma_{\Delta N \to N N} / d \Omega$ 
for the absorption channel.
Conversely, if the partial lifetime is changed then the cross section has 
to be changed accordingly.
Notice that the probabilities of the processes
with the $\Delta$ resonance in the {\it final} state, like e.g. 
$\pi N \to \Delta$ and $N N \to N \Delta$, are {\it not } modified.
We now assume that the overall lifetime is given by Eq. (\ref{taulife})
and define a modified total width 
$\tilde{\Gamma} \equiv \tau^{-1}$. $\tilde{\Gamma}$ can be 
decomposed into modified partial widths $\tilde{\Gamma}_i$:
$\tilde{\Gamma} = \sum_i\,\tilde{\Gamma}_i$. There is one important
aspect: the modified branching ratios, $\tilde{\Gamma}_i/\tilde{\Gamma}$, 
have to be the same as the original ones, $\Gamma_i/\Gamma$,
i.e. $\tilde{\Gamma}_i = \Gamma_i \tilde{\Gamma}/\Gamma 
= \Gamma_i (\Gamma\tau)^{-1}$. This ensures that the measurable
cross sections for multistep processes are correct. To see this we
consider the process $N N \to N N \pi$ with a $\Delta$ 
as an intermediate state. For the cross section one obtains
$\sigma_{N N \to N N \pi} = \sigma_{N N \to N \Delta} B_{\rm out}$,
where $\sigma_{N N \to N \Delta}$ is the production cross section of 
the $\Delta$ resonance in a nucleon-nucleon collision and
$B_{\rm out} = \Gamma_{\Delta \to N \pi}/\Gamma \stackrel{!}{=} 
\tilde{\Gamma}_{\Delta \to N \pi}/\tilde{\Gamma}$ 
is the outgoing branching ratio. Keeping the branching ratios constant,
both quantities, $\sigma_{N N \to N \Delta}$ and $B_{\rm out}$ are not 
modified and hence the (experimentally measurable) cross section 
$\sigma_{N N \to N N \pi}$ is not modified as well.

In the transport simulation the modified partial widths must be treated by 
the Monte-Carlo method in the same way as the usual partial widths. For the 
$\Delta$ resonance, therefore, the probabilities of processes where
the $\Delta$ resonance is present in the {\it initial} state are {\it all} 
multiplied by the same factor $\tilde{\Gamma}/\Gamma=(\Gamma\tau)^{-1}$
to keep the branching ratios constant 
(cf. also Ref. \cite{leupold01} for a different derivation):
\begin{eqnarray}
   \tilde{\Gamma}_{\Delta \to N\pi} = 
                 \Gamma_{\Delta \to N\pi}
                                    (\Gamma\tau)^{-1}~,    \label{gammadec} \\
   {d \tilde{\sigma}_{\Delta N \to N N} \over d \Omega} 
 = {d \sigma_{\Delta N \to N N} \over d \Omega} 
                                    (\Gamma\tau)^{-1}~,    \label{sigmod} \\
   {d \tilde{\sigma}_{\Delta N \to \Delta N} \over d \Omega d M^2} 
 = {d \sigma_{\Delta N \to \Delta N} \over d \Omega d M^2} 
                                    (\Gamma\tau)^{-1}~,    \label{sigmod1} \\
   \tilde{\Gamma}_{\Delta NN \to NNN}
 = \Gamma_{\Delta NN \to NNN} (\Gamma\tau)^{-1}~,  \label{gam3b}
\end{eqnarray}
where $\Gamma_{\Delta \to N\pi}$ is the standard $\Delta$ decay width
in nuclear matter taking into account the Pauli blocking for the final nucleon,
$d \sigma_{\Delta N \to N N} / d \Omega$ is the usual 
differential cross section obtained directly as 
\begin{equation}
{d \sigma_{\Delta N \to N N} \over d \Omega} =
 {1 \over 64 \pi^2} \overline{|{\cal M}|^2}
{p_{NN} \over p_{N\Delta} s} \times {4 \over C_{NN}}~,
                                                             \label{sigabs}
\end{equation}
where $p_{NN}$ and $p_{N\Delta}$ are the c.m. momenta of incoming and
outgoing particles respectively, $s$ is the c.m. energy squared,
$\overline{|{\cal M}|^2}$ is the spin-averaged matrix element squared
given by the one-pion exchange model \cite{DSG86}, 
and  $C_{NN}=2$ (1) if the final nucleons are identical (different).
Analogous modifications have also to be done with the $\Delta$ 
rescattering cross section (\ref{sigmod1}) and 
with the three-body absorption width (\ref{gam3b}), whenever the process 
$\Delta N N \to N N N$ is included in the BUU theory. 

Before discussing BUU results we will specify the explicit forms
of the $\Delta$ resonance width and lifetime.
It has been shown in \cite{Eh93}, that the total width of 
the $\Delta$ resonance in nuclear matter is very close to the vacuum decay 
width. Thus, we take
\begin{equation}
       \Gamma \simeq \Gamma_{\Delta \to N \pi}^{vac}(E_{c.m.}) 
       = \Gamma_0\left({q \over q_0}\right)^3
       {m_\Delta \over E_{c.m.}} {\beta_0^2+q_0^2 \over \beta_0^2+q^2}~,
                                                             \label{gamma1} 
\end{equation}
where the parameterization from \cite{effe1} is used for 
$\Gamma_{\Delta \to N \pi}^{vac}$, $q(E_{c.m.})$ being the pion momentum 
in the rest frame of $\Delta$, $q_0 \equiv q(m_\Delta)$, $\Gamma_0=0.118$
GeV and $\beta_0=0.2$ GeV. 
For simplicity we will also neglect the energy and momentum 
dependence of ${\rm Re}\,\Sigma^+$ putting 
$E_R = {\rm const} = m_\Delta$. In this case the lifetime of Eq. 
(\ref{taulife}) reduces to (\ref{taudel}) with $\delta$ given by
(\ref{phshift}).

Fig. 1 shows the dependence of the $\Delta$ resonance lifetime on the
$\pi N$ c.m. energy as given by the derivative of the phase shift (solid
line) and by the inverse width (dashed line). The shape of the
function $\tau(E_{c.m.})$ is basically dominated by the presence of
the spectral function $A$ in Eq.(\ref{taulife}).
This implies that resonances with mass near the pole mass $m_\Delta$ have
the longest lifetime.

\section{ BUU results } 

In the following we will address the question of whether the difference 
between the lifetime prescriptions could be visible in observable 
quantities. The numerical calculations have been performed on the basis 
of the BUU code in the version of Ref. \cite{effe1} using the soft 
momentum-dependent mean field (SM) with the incompressibility $K=220$ MeV. 
The resonance production/absorption quenching \cite{LCLM01} has been 
implemented in order to reproduce the experimentally measured pion 
multiplicity in central Au+Au collisions at 1 A GeV.   

First, we have performed a BUU calculation of the pion absorption 
on nuclei. To this aim we have selected the experimental data from Ref. 
\cite{Ash81} on reactions $\pi^+$ + C and $\pi^+$ + Fe at the pion beam 
energies $E_\pi =$ 85, 125, 165, 205, 245 and 315 MeV. Fig. 2 shows the 
calculated excitation function of the $\pi^+$ absorption cross section 
in comparison with the data. The standard BUU calculation (dashed lines) 
employing the $\Delta$ lifetime $1/\Gamma$ underpredicts the data at the 
lower energies. Using the lifetime of Eq.(\ref{taulife}) (dotted lines) 
improves the agreement amplifying the absorption peak at $E_\pi=150\div200$ 
MeV. The peak corresponds to the beam energy 
$E_\pi = (m_\Delta^2-m_\pi^2-m_N^2)/2m_N - m_\pi = 192$ MeV at which the 
$\Delta$ resonance is excited at the pole mass. However, the peak position 
is slightly changed by the Fermi motion. The physical reason for the 
increased absorption with the lifetime of Eq.(\ref{taulife}) lies in longer 
living $\Delta$ resonances near the pole mass (see Fig. 1, solid line). This 
increases the probability that a $\Delta$ will be absorbed in the collision 
with a nucleon. Applying now Eq.(\ref{taulife}) for the lifetime and, in 
addition, modifying the $\Delta N \to N N$ and $\Delta N \to \Delta N$ cross 
sections according to Eqs.(\ref{sigmod}),(\ref{sigmod1}) results in the 
absorption cross section (solid lines) practically indistinguishable from the 
standard calculation, since the factor $(\Gamma\tau)^{-1}$ in 
Eq.(\ref{sigmod}) is less than 1 near the pole mass of the $\Delta$.

The discrepancy between our calculations and the data at $E_{\pi} < 250$ 
MeV can be explained by missing the three-body absorption mechanism of 
the $\Delta$ resonance: $\Delta N N \to N N N$. We have performed
the calculation including the three-body absorption as parameterized by
Oset and Salcedo \cite{OS87} (see dash-dotted lines in Fig. 2).
In this case we show only results with the lifetime of Eq.(\ref{taulife})
and the modified absorption and rescattering probabilities in the processes 
$\Delta N \to N N$, $\Delta N N \to N N N$ and $\Delta N \to \Delta N$ 
(Eqs.(\ref{sigmod})-(\ref{gam3b})), since these modifications, when done all 
simultaneously, have only a very small effect on the absorption cross section 
(c.f. solid and dashed lines). We see now a good description at the lower 
energies, but at higher energies the data are overpredicted somewhat.

The authors of Ref. \cite{David99} have observed an influence of the 
$\Delta$ lifetime variations on the $K^+$ in-plane flow in central 
Ni+Ni collisions at 1.93 A GeV. We have studied the $\pi^+$ in-plane flow 
for the system Au+Au at 1 A GeV and b=6 fm. Fig. 3 shows our calculations
in comparison with the data from Ref. \cite{Kint97}. The acceptance of the 
detector \cite{Kint97} for $\pi^+$'s is good only at positive c.m. rapidities.
This causes the measured $<p_x>(Y^{(0)})$-dependence to be asymmetric with 
respect to $Y^{(0)}=0$. At $Y^{(0)}>0$ all calculations generally agree with 
data within errorbars. However, there is a difference in the flow 
($\equiv d<p_x>/dY^{(0)}$ at $Y^{(0)}=0$) between the calculations: the 
calculation with the modified lifetime (\ref{taulife}) and the modified 
cross sections (\ref{sigmod}),(\ref{sigmod1}) produces less (negative) flow 
than the standard calculation, while the results with only modified lifetime 
(\ref{taulife}) are practically the same within statistics with standard ones.
Indeed, $\pi^+$'s exhibit the antiflow due to the superposition of the 
shadowing and the Coulomb repulsion from protons.  Smaller cross sections 
$\tilde{\sigma}_{\Delta N \to N N}$, $\tilde{\sigma}_{\Delta N \to \Delta N}$
near the $\Delta$ pole mass give less shadowing and, therefore, less 
antiflow.

Fig. 4 shows the invariant mass spectrum of the correlated proton-pion 
pairs from the central collision Au+Au at 1.06 A GeV in comparison with 
the data from Ref. \cite{Eskef98}. We have extracted the correlated pairs 
by selecting the proton and pion which are emitted from the same resonance 
and did not rescatter afterwards (see Ref. \cite{LCEM00} for the comparison 
of this method with the background subtraction technique). The shape of
the calculated spectrum well agrees with data, but the peak position is
overpredicted by about 50 MeV. We checked, that the peak position is not
changed and the spectrum gets slightly wider when taking into account also 
those pairs, where the nucleon experienced rescattering on other nucleons 
one or two times. The calculations with the modified lifetime (\ref{taulife}) 
(dotted line) and with modifying both lifetime and cross sections
(\ref{sigmod}),(\ref{sigmod1}) (solid line) result in somewhat sharper peaks 
at the invariant mass of 1.2 GeV.
This is due to longer living $\Delta$ resonances near the pole
mass, which reach the late freeze-out stage and then decay to the 
proton-pion pairs. The $\Delta$ resonances propagate now in an expanding
nuclear matter and, therefore, their absorption is not so effective
as in the pion-nucleus reactions. Thus, increasing the lifetime of the 
$\Delta$ resonances near the pole mass does not lead to their increased 
absorption, unlike the pion-nucleus case.

\section{ Conclusions }

We have performed a comparative study of the two choices
of the $\Delta$ resonance lifetime. Using the inverse width is the 
standard one. However, this choice is physically {\it wrong},
since the resonances propagated in BUU are not asymptotical plane
waves, but intermediate states of the $\pi N$ scattering. The
{\it correct} choice of the resonance lifetime in BUU is the derivative
of the phase shift (c.f. Eq.(\ref{taudel}) or Eq.(\ref{taulife})), which
is the delay time of the scattered wave. A consistent transport theory must 
include also the corresponding modifications of the $\Delta$ resonance
absorption and rescattering probabilities according to 
Eqs.(\ref{sigmod})-(\ref{gam3b}). Surprisingly, we have found a quite weak
sensitivity of the calculated pionic observables on the lifetime 
prescription, once the cross sections are modified accordingly, despite
of the strong difference in the c.m. energy dependence of both
lifetimes (Fig. 1).
Nevertheless, detailed comparison with more precise data on the pion flow 
and $(p,\pi)$ correlations is necessary to confirm the correct
choice of the $\Delta$ lifetime in nuclear matter. 

\begin{acknowledgments}
Authors are grateful to W. Cassing for useful discussions.
This work was supported by GSI Darmstadt.
\end{acknowledgments}

\clearpage

\thispagestyle{empty}

\begin{figure}

\includegraphics{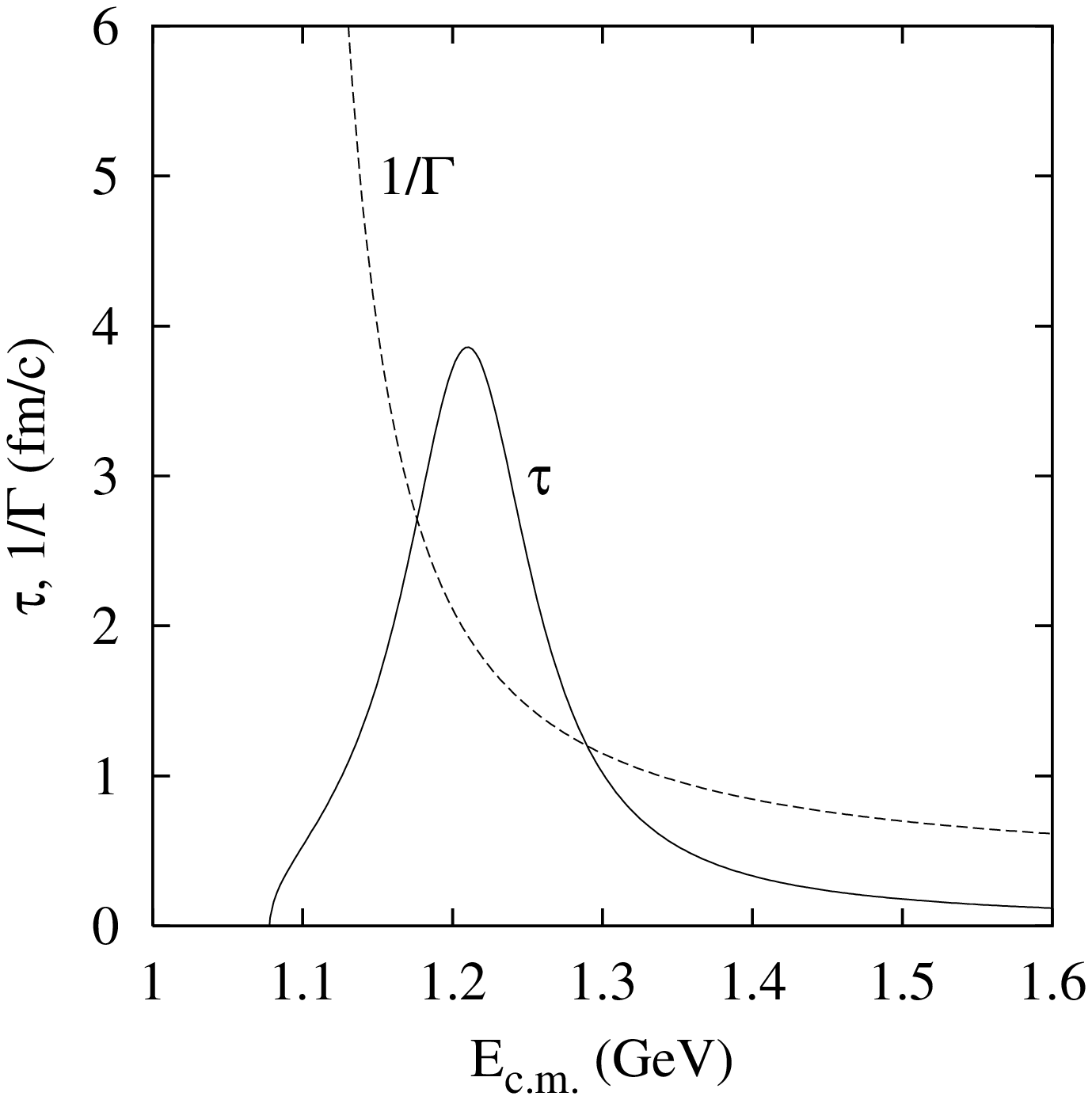}

\caption{The inverse width $1/\Gamma$ (dashed line), where 
$\Gamma$ is given by Eq.(\ref{gamma1}), and the lifetime
(solid line, Eq.(\ref{taulife})) of the $\Delta$ resonance
as functions of the total c.m. energy of the pion and nucleon.}

\end{figure}

\clearpage

\thispagestyle{empty}

\begin{figure}

\includegraphics{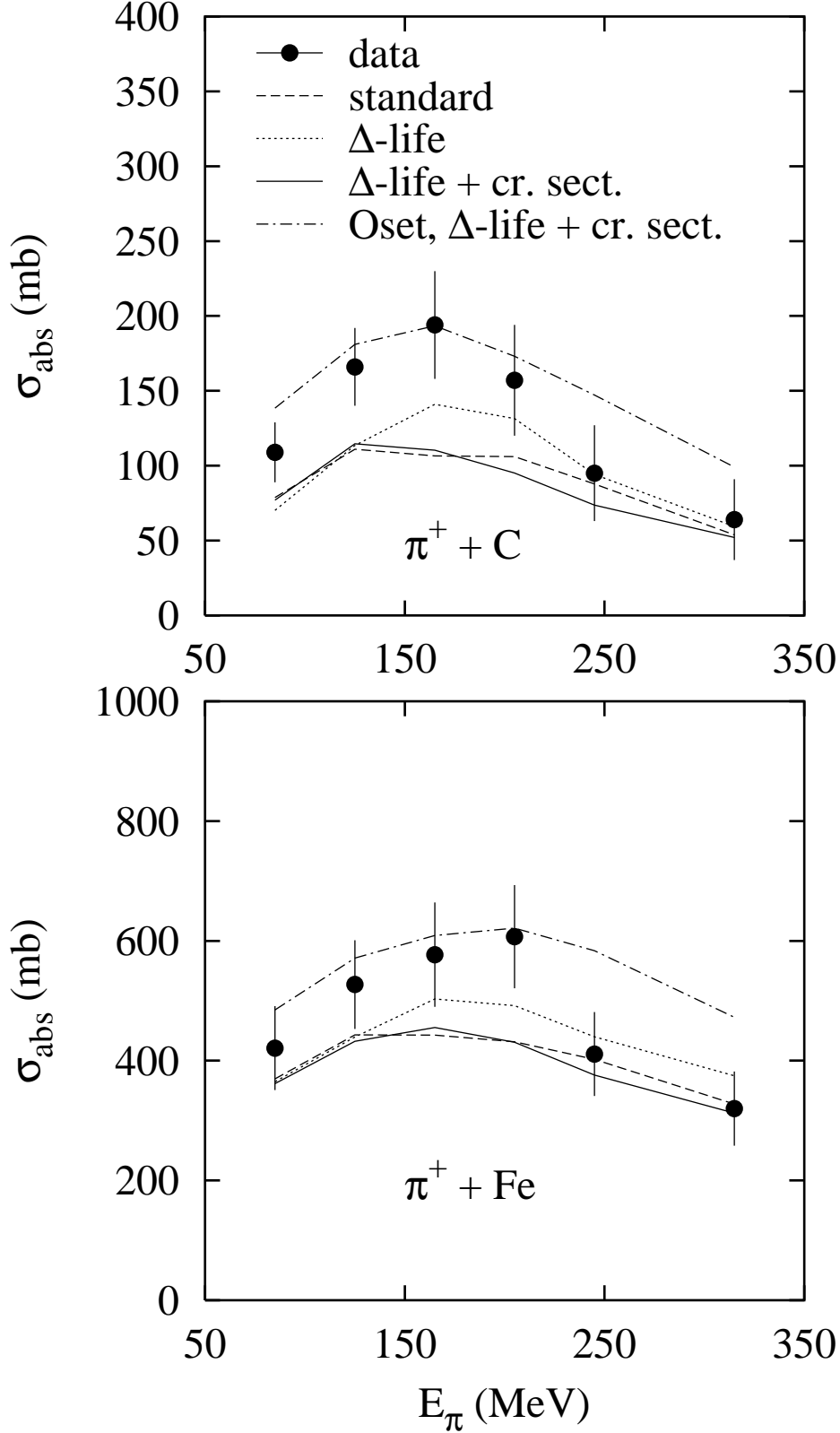}

\caption{The beam energy dependence of the $\pi^+$ absorption
cross section on carbon (upper panel) and on iron (lower panel).
Standard BUU calculations (with $1/\Gamma$ $\Delta$ lifetime) 
are represented by dashed line, while dotted and solid lines show 
respectively the results with the modified lifetime of Eq.(\ref{taulife})
and with both modified lifetime and cross sections
$\Delta N \to N N$, $\Delta N \to \Delta N$ of Eqs.(\ref{sigmod}),
(\ref{sigmod1}). Dash-dotted line shows
the calculations including the three-body absorption contribution 
\cite{OS87} (see text for details). Experimental data are from Ref. 
\cite{Ash81}.}

\end{figure}

\clearpage

\thispagestyle{empty}

\begin{figure}

\includegraphics{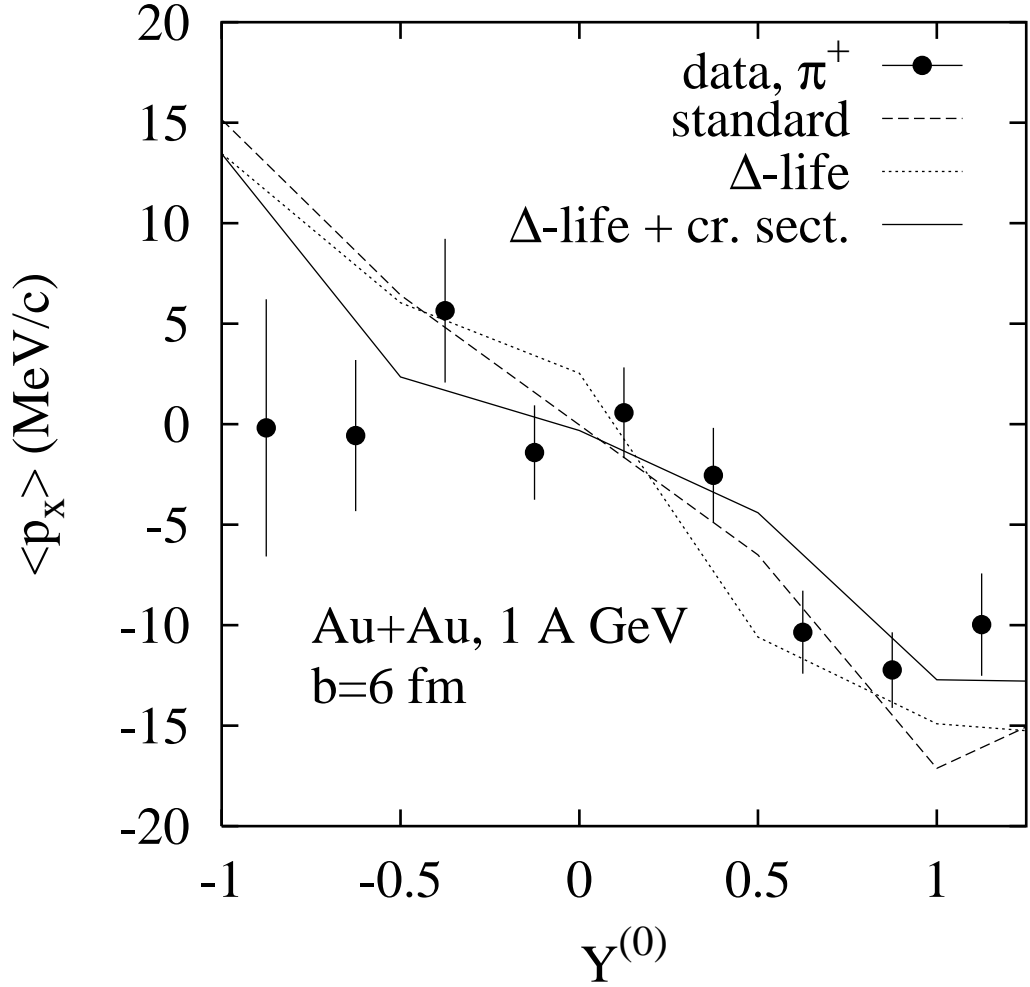}

\caption{The $\pi^+$ average transverse momentum in the reaction plane
vs. normalized c.m. rapidity $Y^{(0)} \equiv (y/y_{proj})_{c.m.}$ for the 
collision Au+Au at 1 A GeV and b=6 fm. Calculated curves are denoted as in 
Fig. 2. Data are from Ref. \cite{Kint97}.}

\end{figure}

\clearpage

\thispagestyle{empty}

\begin{figure}

\includegraphics{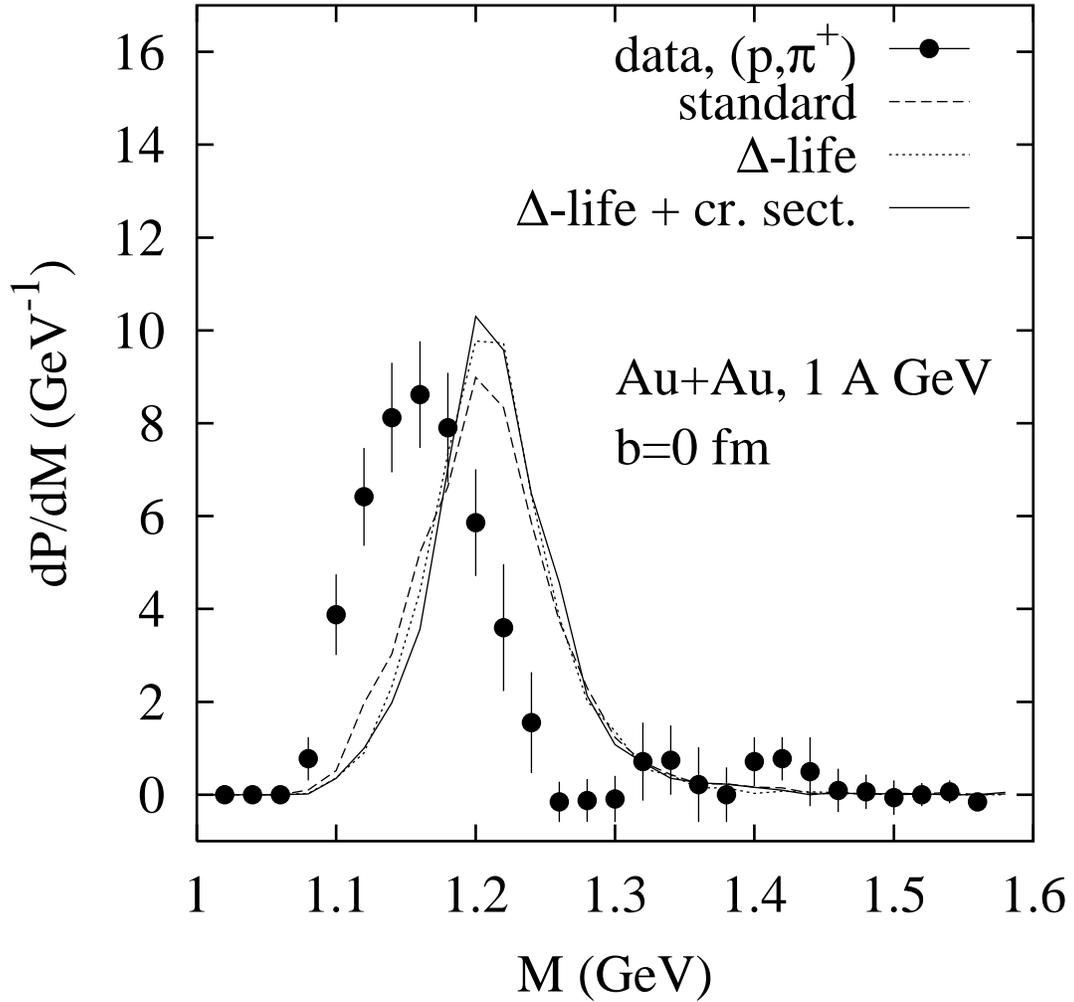}

\caption{Invariant mass spectrum of the correlated ($p,\pi^+$) pairs
for the system Au+Au at 1 A GeV. Calculated curves are denoted as in Fig. 2.
Data are from Ref. \cite{Eskef98}.}

\end{figure}

\end{document}